\documentclass[epj]{svjour}
\usepackage{epsfig}

\begin{document}
\title{Optical measurements of the superconducting gap in MgB$_2$}
\author{B. Gorshunov\inst{1} \and C. A. Kuntscher\inst{1} \and P.
Haas\inst{1} \and M. Dressel\inst{1}\thanks{email:
dressel@pi1.physik.uni- stuttgart.de}
 \and F. P. Mena\inst{2} \and A. B. Kuz'menko\inst{2}
\and D. van der Marel\inst{2} \and T. Muranaka\inst{3} \and J.
Akimitsu\inst{3} }
\institute{1. Physikalisches Institut, Universit\"at Stuttgart,
Pfaffenwaldring 57, D-70550 Stuttgart, Germany \and Laboratory of
Solid State Physics, Materials Science Centre, Nijenborgh 4, 9747
AG Groningen, The Netherlands \and Department of Physics,
Aoyama-Gakuin University, 6-16-1 Chitsedai, Setagaya-ku, Tokyo
157, Japan}
\date{Received: date 7 March 2001 / Revised version: \today}
%
\abstract{Far-infrared reflectivity studies on the polycrystalline
intermetallic compound MgB$_2$ with a superconducting transition
temperature $T_c=39$~K were performed at temperatures 20~K to
300~K. We observe a significant raise of the
superconducting-to-normal state reflectivity ratio below $70~{\rm
cm}^{-1}$, with a maximum at about $25-30~{\rm cm}^{-1}$, which
gives a lower estimate of the superconducting gap of
$2\Delta(0)$$\approx3-4$~meV.
} 
\maketitle

Recently superconductivity in the binary intermetallic compound
MgB$_2$ with a $T_c$ close to 40~K was reported by Akimitsu {\em
et al.} \cite{Akimitsu01}. Most studies performed on this
compound yet indicate that MgB$_2$ consistently behaves as a
phonon mediated superconductor within the framework of the BCS
theory, probably in a strong coupling limit
\cite{Budko01,Finnemore01,Jung01,Kotegawa01}. Up to now no
optical investigations of this material were reported, and this
is in part due to the poor quality of the presently available
samples, which are sintered polycrystals. It is well known,
however, that optical spectroscopy is an extremely powerful
method for studying the materials in the superconducting (SC)
state and is able of providing information on such important
parameters as the SC energy gap, penetration depth, coherence
effects,  scattering mechanism, etc. \cite{Tinkham96,Timusk89}.

In this short note we report on far-infrared reflectivity
measurements performed on a sintered pellet of MgB$_2$. We are
fully aware that these experiments bear inherent problems, mainly
connected with the surface roughness, which we cannot overcome at
this point. Nevertheless, we have attempted to extract important
information on the intrinsic properties of MgB$_2$ from our
reflectivity data.

\begin{figure}[b]
\centerline{\psfig{file=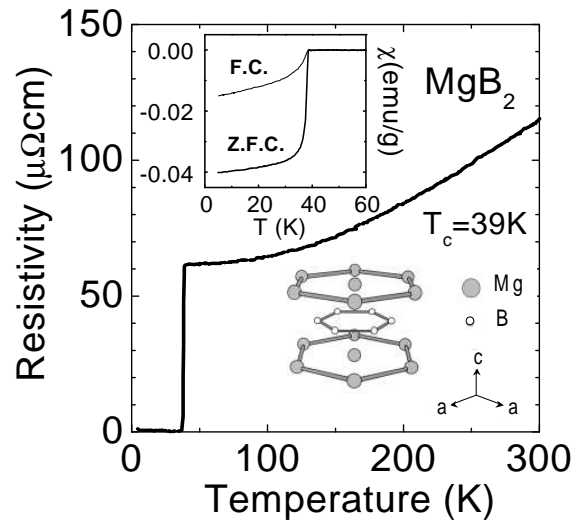,height=7cm,clip=}}
\caption{\label{fig:dc_res} Temperature dependence of the dc
resistivity and of the magnetic susceptibility (field-cooled,
zero-field-cooled) of MgB$_2$ and the unit cell of MgB$_2$.}
\end{figure}

MgB$_{2}$ has a hexagonal structure with $P6/mmm$ symmetry. It
crystallizes in the so-called AlB$_{2}$ structure where the boron
atoms are located at a primitive honeycomb lattice, consisting of
graphite-type sheets. The structure of MgB$_2$ is shown in
Fig.~\ref{fig:dc_res}; the dimensions of the unit cell are
$a$=3.086~\AA\ and $c$=3.524~\AA, according to our X-ray
analysis. The borons span hexagonal prisms; the large, almost
spherical pores are filled by Mg which acts as a spacer. Similar
to graphite, the distance between the boron planes is larger by a
factor of two compared to the intraplanar B-B bonds, and hence
the B-B bonding is strongly anisotropic (two-dimensional).

Powder of high purity MgB$_2$ was pressed in a pellet and treated
in Ar atmosphere at $\approx$900$^\circ$C; details on the sample
preparation are reported in Ref.~\cite{Akimitsu01}. The quality
of the sintered samples was checked by X-ray analysis,
resistivity, and susceptibility measurements. In
Fig.~\ref{fig:dc_res} the temperature dependence of the dc
resistivity and of the magnetic susceptibility are plotted. The
SC phase transition is characterized by a width of less than 1 K
from resistivity and of around 5 K from the susceptibility
measurements. At $T$=300~K we find $\rho$=$115~\mu\Omega$cm, and
just above the SC transition temperature  $T_c$=39~K the
resistivity has decreased by a factor of 2. The resistivity in
the normal state can best be fitted by a power-law temperature
dependence $T^{\alpha}$ with $\alpha$=2.5 up to 250~K; behaviors
with $\alpha$=2 and 3 were reported earlier
\cite{Finnemore01,Jung01}. The susceptibility (inset of Fig.1) is
decreasing rapidly in the temperature range 39 - 34~K by 0.034
emu/g, whereas for lower temperatures it decreases only slowly to
the value -0.04 emu/g. Magnetization measurements allow us to
estimate a SC fraction of more than 50\%.

For the optical investigations the sintered pellet of MgB$_2$ was
cut to a piece of 5$\times$5$\times$2~mm$^3$ size and polished. We
note that it had a smooth but not shiny surface, with a remaining
roughness due to pores. The scanning electron microscopy analysis
revealed some traces of oxygen on the sample surface which may
correspond to a layer of the MgO on it.

Here we report the far-infrared reflectivity spectra measured in
the grazing (80 degrees) incidence geometry using a FT-IR
spectrometer. The grazing incidence geometry has been chosen
because in this case the reflectivity is more sensitive to
changes of optical conductivity due to opening of the
superconducting gap. This experimental technique has been
previously used to measure the $s$-wave gap of NbN and the
$d$-wave gap of La$_{1.85}$Sr$_{0.15}$CuO$_{4}$
~\cite{Schutzmann97}. In addition, normal incidence spectra were
collected in the same frequency range, which were qualitatively
consistent with the grazing reflectance data reported in this
paper.

In our measurements, we have observed that the copper block on
which the sample is mounted in the cryostat and the sample surface
had quite different temperatures. In order to get a precise
surface temperature, an additional Pt thermoresistor was later
attached to the sample face. It turned out that the temperature
difference is strongly enhanced below 40 K, suggesting that the
large thermal gradient is due to strong lowering of the sample
thermal conductivity in the SC state. In particular, the lowest
surface temperature we could achieve was only about 20 K, while
the cryostat cold finger was at liquid helium temperature. This
additional temperature calibration has been used in the analysis
presented in this paper. However, we should note that the
systematic error of this measurement on this rather thick sample
is about 2-3 K.

\begin{figure}
\centerline{\psfig{file=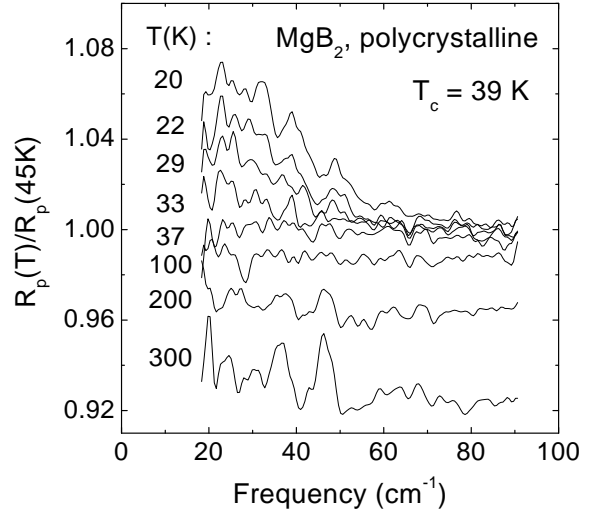,height=7cm,clip=}}
\caption{\label{fig:Ratio} Frequency dependence of the ratio of
the reflectivity of MgB$_2$ measured in grazing incidence
geometry ($p$ - polarization) at various temperatures $T$ (above
and below $T_c$) to the gra\-zing incidence reflectivity at
$T$=45~K.}
\end{figure}

\begin{figure}
\centerline{\psfig{file=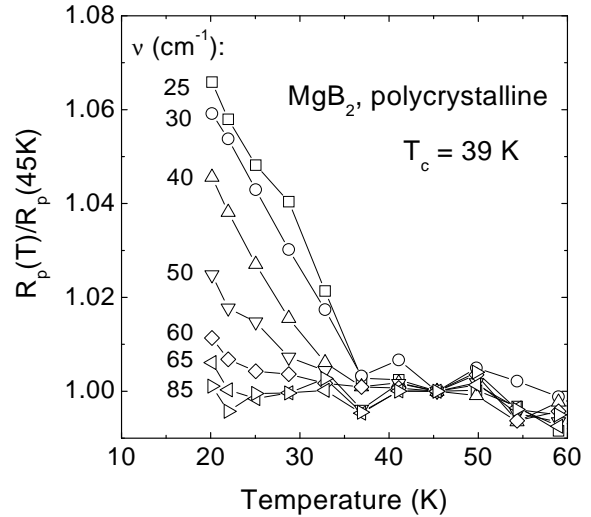,height=7cm,clip=}}
\caption{\label{fig:Reflectivity} Temperature dependence of the
grazing incidence ($p$-polarization) reflectivity ratio of
MgB$_2$.}
\end{figure}

Fig.~\ref{fig:Ratio} presents the intensity reflected from the
sample at various temperatures normalized to the intensity
reflected at 45 K, i.~e., slightly above $T_c$. We clearly see a
rise of the reflectivity ratio (RR) above 1, starting below
$T_c$. This is seen also in Fig.~\ref{fig:Reflectivity}, where the
temperature dependences of the RR is plotted for fixed
frequencies. The frequency below which the RR starts to increase
above 1 shifts to higher values with decreasing $T$; for the
lowest temperature $T$=20~K the increase starts at around
$\nu$=70~cm$^{-1}$. As expected, the RR spectra in the SC state
reveal maxima, meaning
that RR=1 at $\nu$=0. %

The observed variation of the reflectivity is reminiscent of the
behavior found in conventional superconductors and, to a certain
extent, in the high-temperature cuprates.  There, the
condensation of carriers into pairs and the opening of the energy
gap $2\Delta$ in the density of states reveal themselves as a
decrease of the conductivity starting around the corresponding
frequency $\nu$=$2\Delta/hc$, and as the response in the
dielectric constant of $\epsilon\propto -1/\nu^2$, both leading to
pronounced changes of reflectivity and a maximum in the RR at
around $\nu$=$2\Delta/hc$ \cite{Tinkham96,Timusk89}. Accordingly,
we associate the changes we observe for MgB$_2$ by entering the
SC state with the response of condensed superconducting pairs,
which is only in part obscured by the surface scattering. (We
note that the rise in reflection for an essentially two
dimensional system is not always an indication of the SC gap; for
instance, the gap-like feature in the reflectivity spectrum of
polycrystalline La$_{1.85}$Sr$_{0.15}$CuO$_{4}$ \cite{Sherwin88}
was later explained by the manifestation of the $c$-axis
Josephson plasmon \cite{Tamasaki92}).

A straight-forward assignment based on an isotropic $s$-wave gap
would associate the gap energy $2\Delta$ with the frequency where
the RR reaches its maximum (around $25-30~{\rm cm}^{-1}$,
according to Fig.2), providing an unrealistically small estimate
of $2\Delta$ = 3 - 4 meV. On the other hand it has been
anticipated that superconducting gap is {\underline {not}}
uniform and has different values for different four sheets of the
Fermi surface \cite{Liu01}. With this interpretation the RR
maximum should correspond to the minimum value of the anisotropic
gap. This is also consistent with a large variation of $2\Delta$
values obtained from tunneling, Raman, photoemission, and NMR
techniques
\cite{Kotegawa01,Rubio01,Sharoni01,Kaparetrov01,Schmidt01,Takahashi01,Chen01}
ranging from 4 to 17~meV

In conclusion, we have investigated the optical properties of
sintered MgB$_2$ in the normal and superconducting state by
measuring the ratio of the reflected intensity in the
superconduc\-ting state to that in the normal state. An increase
of the ratio in the superconducting state is reliably detected
and interpreted as the signature of the superconducting gap with
the lowest value of $2\Delta(0)$$\approx3-4$~meV.


\begin{thebibliography}{99}
\bibitem{Akimitsu01}J.
J. Akimitsu, Symposium on Transition Metal Oxides, Sendai, January
10, 2001; J. Nagamatsu {\it et al.}, Nature {\bf 410}, 63 (2001).

\bibitem{Budko01}
S. L. Bud'ko {\it et al.}, Phys.\ Rev.\ Lett.\ {\bf 86}, 1877
(2001).

\bibitem{Finnemore01}
D. K. Finnemore {\it et al.}, Phys.\ Lett.\ B {\bf 86}, 2420
(2001).

\bibitem{Jung01}
C. U. Jung {\it et al.}, cond-mat/0102215.

\bibitem{Kotegawa01}
H. Kotegawa {\it et al.}, cond-mat/0102334.


\bibitem{Tinkham96}
M. Tinkham, {\em Introduction to Superconductivity}, 2nd edition
(Mc Graw-Hill, New York, 1996).

\bibitem{Timusk89}
T. Timusk, D. Tanner, in: {\em Physical Properies of High
Temperature Superconductors I}, ed. by D. M. Ginsberg (World
Scientific, Singapore, 1989); D. Tanner, T. Timusk, in: {\em
Physical Properies of High Temperature Superconductors III}, ed.
by D. M. Ginsberg  (World Scientific, Singapore, 1992).

\bibitem{Sherwin88}
M. S. Sherwin {\it et al.}, Phys.\ Rev.\ B {\bf 37}, 1587 (1988).

\bibitem{Tamasaki92}
K. Tamasaki {\it et al.}, Phys.\ Rev.\ Lett.\ {\bf 69}, 1445
(1992).


\bibitem{Schutzmann97}
H.S. Somal {\it et al.}, Phys.\ Rev.\ Lett.\ {\bf 76}, 1525
(1996).

\bibitem{Liu01}
A.Y. Liu {\it et al.}, cond-mat/0103570.

\bibitem{Rubio01}
G. Rubio-Bollinger {\it et al.}, cond-mat/0102242.

\bibitem{Sharoni01}
A. Sharoni, {\it et al.},  cond-mat/0102325.

\bibitem{Kaparetrov01}
G. Karapetrov {\it et al.}, cond-mat/0102312.

\bibitem{Schmidt01}
H. Schmidt {\it et al.}, cond-mat/0102389.

\bibitem{Takahashi01}
T. Takahashi {\it et al.}, cond-mat/0103079.

\bibitem{Chen01}
X.K. Chen {\it et al.}, cond-mat/0104005.

\end{thebibliography}
\end{document}